\documentclass[fleqn,twoside]{article}
\usepackage{espcrc2}
\usepackage[dvips]{graphics}

\begin{document}

\title{Hadron spectroscopy of twisted mass lattice QCD at $\beta$ = 6.0.
\thanks{Presented by C.\ McNeile.}}

\author{UKQCD Collaboration, Craig~McNeile
and  Chris~Michael,\address{Department of Mathematical Sciences, 
University of\ Liverpool, L69 3BX, UK}
}

\begin{abstract}
Simulations that use the clover action in quenched QCD calculations
have a lower limit to the quark mass that can be reached, because of
the fluctuations caused by exceptional configurations.  From this low
statistics study, we find that the twisted clover action, recently
introduced by the ALPHA collaboration, can be used to simulate
quenched QCD at quark masses below those attainable by simulations that use 
the clover action.
\end{abstract}
 
\maketitle

\section{INTRODUCTION}

The masses of the so called light quarks used in lattice QCD
calculations are not light enough for chiral perturbation theory to be
a reliable guide to extrapolating the results to the physical quark masses.
For example, in our recent work on determining the
parameters of the chiral QCD Lagrangian~\cite{Irving:2001vy}, we were
unable to test fully the chiral perturbation theory predictions
because the dynamical quarks were too heavy and there was a lower
limit on the mass of the valence quarks caused by exceptional
configurations.

The ALPHA collaboration have developed~\cite{Frezzotti:2000nk} the
twisted clover fermion action, specifically to study the light quark
mass region of QCD. The ALPHA collaboration~\cite{DellaMorte:2001ys}
have used the twisted clover action in numerical simulations within
the Schr\"{o}dinger functional formalism.
We report the first computation of hadron spectroscopy using the
twisted clover fermion operator with periodic in space and anti-periodic 
in time boundary conditions.

\section{THE TWISTED-FERMION OPERATOR}

\begin{equation}
S_F = \sum_{x}
\overline{\psi} (x) 
(D + m_q + i \mu_q \gamma_5 \tau_3) \psi(x)
\label{eq:twistaction}
\end{equation}
The operator $D$ is the fermion operator
that is independent of the quark mass.
The mass term 
\begin{equation}
i \mu_q \gamma_5 \tau_3  = i \mu_q \gamma_5\left(
\begin{array}{cc}
1 & 0 \\
0 & -1 \\
\end{array}
\right)
\end{equation}
has no additive renormalisation, hence it provides
a lower limit to the eigenvalue spectrum of the fermion
operator. 

The main disadvantages of the twisted clover action are
that parity and flavour symmetry are explicitly broken.
However, the flavour symmetry breaking looks very different to
the mechanism in the staggered fermion 
formulation~\cite{Lepage:1998vj},
so the effects of the 
flavour symmetry breaking may be less of a problem.

The transformation
\begin{equation}
\psi' \rightarrow   e^{i \alpha \gamma_5 \tau_3/2}  \psi 
 \; \; \; \; \; \; \; \;
\overline{\psi'} \rightarrow \overline{\psi } 
e^{i \alpha \gamma_5 \tau_3/2}   
\label{eq:magicTRANS}
\end{equation}
transforms the action in Eq.~\ref{eq:twistaction},
when D anti-commutes with $\gamma_5$ as in the continuum,
into the standard QCD action
\begin{equation}
S_F = \sum_{x}
\overline{\psi'} (x) 
(D +  M_Q) \psi'(x)
\label{eq:backTOqcd}
\end{equation}
where
\begin{eqnarray}
M_Q & = & \sqrt{m_q^2 + \mu_q^2 } \nonumber \\
\tan \alpha &= & \frac{ \mu_q } { m_q  }  \nonumber
\end{eqnarray}
The ALPHA collaboration~\cite{Frezzotti:2000nk,Frezzotti:2001ea}
advocate the following 
procedure. The simulations are carried out using the 
twisted clover operator on the lattice. The results are
then matched to the continuum twisted fermion operator.
The rotations in Eq.~\ref{eq:magicTRANS} are then used to
rotate back to the standard QCD fermion action in the continuum.
The rotations should reduce the effect of the flavour and 
parity symmetry breaking on the physical spectrum.
The connection between QCD and twisted QCD
could be more complicated, when there is spontaneous
breakdown of parity and flavour symmetry, such as 
in the Aoki phase~\cite{Sharpe:1998xm}.

Another way to look at the connection between the twisted and the
traditional clover action is to think that the rotations
Eq.~\ref{eq:magicTRANS} are done on the
lattice~\cite{Schierholz:1998bq}. In this viewpoint the rotations
(Eq.~\ref{eq:magicTRANS}) are similar to the effect of a preconditioner,
except that the connection between the twisted and standard clover
actions is not exact, because the transformation in
Eq.~\ref{eq:magicTRANS} also rotates the Wilson term in the operator,

The formalism for the $O(a)$ improvement of the twisted fermion
operator has been described in~\cite{Frezzotti:2001ea}. 

\section{INTERPOLATING OPERATORS}

The parity and flavour symmetry violating parts of the twisted clover
action make the construction of the interpolating operators
nontrivial. ALPHA~\cite{Frezzotti:2000nk} use the 
twisting rotations in Eq.~\ref{eq:magicTRANS}
to construct the interpolating operators.

Consider a general meson interpolating operator
\begin{equation}
\overline{\psi} \Gamma \frac{\tau_i}{2} \psi
\end{equation}
For the Wilson/clover action all three meson
operators (i = 1,2,3) are equivalent. The flavour symmetry 
breaking term in the twisted action breaks
the equivalence~\cite{Sharpe:1998xm}.

For i =1,2 the generic meson operators are
\begin{eqnarray}
\overline{u'} \Gamma  d' &= & 
\overline{u} (\cos(\alpha /2 ) +  i \gamma_5 \sin(\alpha/2) )
\nonumber  \\ 
& \Gamma &
(\cos(\frac{\alpha}{2} ) -  i \gamma_5 \sin(\frac{\alpha}{2}) )
d
\end{eqnarray}
For $\Gamma$ = 1 (scalar), $\gamma_5$ (pion), $\gamma_{i} \gamma_{j}$
($b_1$) and $\gamma_4 \gamma_i$ (second rho operator).
\begin{equation}
\overline{u'} \Gamma  d' = \overline{u}  \Gamma d
\end{equation}
For $\Gamma$ = $\gamma_4 \gamma_5$ (pion), $\gamma_4$ ("exotic"), 
$\gamma_{i}$ (rho)
\begin{equation}
\overline{u'} \Gamma  d'  =  \cos(\alpha) \overline{u} \Gamma d 
 +  i \sin(\alpha) \overline{u} \gamma_5 \Gamma d
\end{equation}

The neutral meson operator is proportional to
\begin{equation}
\overline{u} \Gamma u - \overline{d} \Gamma d
\end{equation}
The neutral meson operator has the opposite 
connection between the gamma matrix and the 
mixing as for the charged meson operators.
Disconnected fermion loops are required to compute the correlators for
the neutral meson operator, because the twisted fermion 
operator does not obey
$M^{\dagger} = \gamma_5 M \gamma_5$~\cite{Sharpe:1998xm}.

To compute the correlator for the nucleon with 
an interpolating operator, such as:
\begin{equation}
P(x)_i  =  \epsilon_{abc} ( u^a(x)^T C \gamma_5 d^b(x) ) u^c_i(x) ,
\nonumber \\
\end{equation}
separate inversions are required for the up and down twisted
clover propagators. These can be rotated back to the clover fermion
matrix using Eq.~\ref{eq:magicTRANS}, before the nucleon correlator is
constructed.

\section{NUMERICAL RESULTS}

The simulations were performed at $\beta=6.0$ on $16^3 48$ 
lattice in the quenched approximation. 
The numerical value of the clover coefficient
determined by the ALPHA collaboration~\cite{Luscher:1997ug}
was used.
The UKQCD collaboration~\cite{Bowler:1999ae} found, at the same
parameters, evidence for exceptional configurations at $M_{PS} / M_{V}
\sim 0.54$, hence we aimed to use the twisted clover action to 
explore the quark mass region under that limit.
For this exploratory study we only used 30 configurations.

UKQCD found~\cite{Bowler:1999ae} 
$\kappa_{critical} = 0.135252$, so we initially ran at
($\kappa = 0.135$, $\mu$ = 0.02), 
($\kappa = 0.135$, $\mu$ = 0.01), 
and ($\kappa = 0.135$, $\mu$ = 0.005). We were aiming for 
$\alpha \sim \frac{\pi}{2}$, but this $\kappa$ value was not close
enough to $\kappa_{critical}$, and the $\alpha$ values for the
runs $\mu$ = 0.02, 0.01, and 0.005 were: $0.41 \pi$, 
$0.33 \pi$, $0.24 \pi$. For all these runs we used the standard
stabilised biconjugate gradient algorithm. To increase 
$\alpha$ to be close to $\frac{\pi}{2}$ we ran at
($\kappa = 0.13525$ , $\mu$ = 0.01) and 
($\kappa = 0.13525$ , $\mu$ = 0.005). For the inversions
with $\kappa = 0.13525$, the conjugate gradient algorithm had
to be used.

\begin{figure}[htb]
\vspace{-2.7cm}
\scalebox{0.30}{\includegraphics{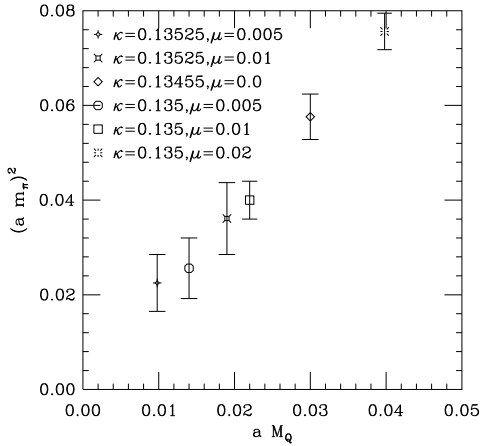}}
\caption{The pion mass 
squared as a function of renormalised
quark mass.}
\label{eq:pionFIG}
\vspace{-1.3cm}
\end{figure}

With the low statistics we could only 
obtain satisfactory fits to
the pion correlator. 
Factorising fits~\cite{Bowler:1999ae} 
were performed to a 2 by 2 matrix, with a basis
of fuzzed and local correlators~\cite{Bowler:1999ae}.
The pion mass as a function of the renormalised 
quark mass ($M_Q$ in Eq.~\ref{eq:backTOqcd}~\cite{Frezzotti:2000nk}, 
with $\mu_q/Z_P$ and 
$m_q= Z_A m_{pcac} /Z_P$, where $m_{pcac}$ is the mass
from the PCAC relation ) is plotted 
in Fig.~\ref{eq:pionFIG}. The pion mass from the twisted 
clover simulation with the 
parameters $\kappa=0.13525, \mu=0.005$ is $\sim 60$  \% lower
than the pion mass  obtained from the clover action result
at the point where exceptionals become a problem
($\kappa = 0.13455$).

\section{ANALYSIS OF THE EIGENVALUES}

It is instructive to consider the effect of the $\gamma_5$ term on the
eigenvalue spectrum of the clover fermion matrix.  Consider the
eigenvalue equation for a single flavour of the twisted clover action.
\begin{equation}
(M + i \mu \gamma_5 ) x(\mu) = \lambda(\mu) x(\mu)
\label{eq:eigEQN}
\end{equation}
where $M$ is the fermion matrix of the 
clover action.
The relation $M^{\dagger} = \gamma_5 M \gamma_5$ 
implies that
the eigenvalues of $M$ are either real, or come in 
complex conjugate pairs (see~\cite{Gattringer:1998ab} for
a review). 

The real eigenvalues of $M$ are also 
eigenvalues of $\gamma_5$, hence the 
eigenvectors of the clover fermion matrix 
with real eigenvalues are also eigenvectors
of the twisted matrix with complex eigenvalues.

The complex eigenvalues of $M$ are not eigenvalues of 
$\gamma_5$, so there is no simple connection between
the eigenvalues of the twisted clover fermion operator and those
of the clover fermion operator.
Perturbation theory shows that 
the first order perturbation~\cite{Golub} 
of the $i \gamma_5 \mu$ term
on the spectrum is 
\begin{equation}
\mid \frac{d \lambda(\mu)}{d \mu} \mid =
\mid  \frac{y^\dagger \gamma_5 x}{y^{\dagger} x} \mid
\end{equation}
where $y$ is the left eigenvector of the clover 
fermion operator $M$.
Unfortunately the symmetries of the clover operator $M$ do not constrain
the matrix element of $y^\dagger \gamma_5 x$. The complex 
eigenvalues of $M$ are changed at $O(\mu)$ by the 
$i \mu \gamma_5$ term.

We are using the ARPACK package to numerically study the 
eigenvalues of the twisted clover action.

\section{CONCLUSIONS}

In the quenched approximation, the twisted clover action has allowed
us to simulate at lighter pion masses than could be reached by the
clover action.  We are increasing the statistics at the lightest quark
mass.
%
%

We thank S. Sint and R. Frezzotti for
discussions.


\end{document}